\documentclass[fleqn,10pt]{wlscirep}
\usepackage{amsmath,amssymb}
\usepackage{mathtools}
\usepackage{xcolor}
\usepackage{hyperref}
\hypersetup{
    pdfcreator={Hello},
    pdfproducer={World}
}

\title{Glass-on-Glass Fabrication of Bottle-Shaped Tunable Micro-Lasers and Their Applications}

\author[1,*]{Jonathan M. Ward}
\author[1,2]{Yong Yang}
\author[1]{S\'{\i}le Nic Chormaic}
\affil[1]{Light-Matter Interactions Unit, Okinawa Institute of Science and Technology Graduate University, Onna-son, Okinawa 904-0495, Japan.}
\affil[2]{National Engineering Laboratory for Fiber Optics Sensing Technology, Wuhan University of Technology, Wuhan, 430070, China.}
\affil[*]{jonathan.ward@oist.jp}

\begin{abstract}
We describe a novel method for making microbottle-shaped lasers by using a CO$_2$ laser to melt Er:Yb glass onto silica microcapillaries or fibres. This is realised by the fact that the two glasses have different melting points. The CO$_2$ laser power is controlled to flow the doped glass around the silica cylinder.  In the case of a capillary, the resulting geometry is a hollow, microbottle-shaped resonator. This is a simple method for fabricating a number of glass WGM lasers with a wide range of sizes on a single, micron-scale structure. The Er:Yb doped glass outer layer is pumped at 980 nm via a tapered optical fibre and whispering gallery mode (WGM) lasing is recorded around 1535 nm. This structure facilitates a new way to thermo-optically tune the microlaser modes by passing gas through the capillary.  The cooling effect of the gas flow shifts the WGMs towards shorter wavelengths, thus thermal tuning of the lasing modes over 70 GHz is achieved. Results are fitted using the theory of hot wire anemometry, allowing the flow rate to be calibrated with a flow sensitivity as high as 100 GHz/sccm. Strain tuning  of the microlaser modes by up to 50 GHz is also demonstrated.
\end{abstract}
\begin{document}

\flushbottom
\maketitle
\thispagestyle{empty}

\section*{Introduction}

The fabrication of glass optical whispering gallery microlasers can be achieved in a limited number of ways. One method involves making a single cavity  by drawing out a glass wire from a piece of doped glass and melting the tip of the wire to form a spherical resonator \cite{Chen2009}. Alternatively, many spherical resonators can be made simultaneously by passing particles of doped glass through a furnace \cite{Lissillour1998, Ward2010fab}. Both of these methods are tedious; in the first case, only one resonator can be made at a time. In the second case, individual spheres must be selected and glued to some other structure, e.g. the tip of a fibre for ease of manipulation.  For both methods, only a single resonator is selected and brought to an evanescent wave guide coupler for optical excitation. On-chip fabrication of a large number of active glass resonators is  possible by doping the chip before fabrication either with ion implantation \cite{Min2004} or by Solgel coating \cite{Yang2005sol,Zheng2015}. For passive devices, individual resonators - such as a microsphere - can also be activated by coating using the Solgel process \cite{Yang2003,Fan2013}.The on-chip method is obviously more complicated and requires much more equipment than the simple heating methods, but it has the advantage  that the coupling between the excitation waveguide and any cavities on the chip can be easily achieved by moving the chip relative to the waveguide. Coating a tapered optical fiber tip with a layer of erbium doped phosphate glass and then melting the tip into a sphere has also been demonstrated to produce lasing microspheres\cite{Dong2008,Dong2010}.  

Here, we present a simple heating method for fabricating a number of glass whispering gallery mode (WGM) lasers with a range of sizes on a single micron-scale structure that can be easily manipulated relative to the excitation waveguide and can, in principle, be packaged onto a millimetre scale chip. We experimentally demonstrate the coating of tapered optical fibres and microcapillaries  with a layer of Er:Yb doped phosphate laser glass. This is achieved by the simple fact that the two glasses have very different melting temperatures, around 1,500$^{\circ}$C for silica and 500$^{\circ}$C for the phosphate glass. The Er:Yb doped glass outer layer is pumped at 980 nm and lasing is observed at 1535 nm.  Microlasers with diameters ranging from 22 $\mu$m to 232 $\mu$m are made on the same structure.

A desired feature of any laser is the ability to tune the frequency of the laser output. Tuning a micron-scale whispering gallery resonator in a fashion that does not add to the footprint of the device or require complicated fabrication is a nontrivial task. The main approaches so far include the use of external heaters \cite{Chiba2004,Suter2007}, application of stress/strain via an external clamp \cite{Klitzing2001,Sumetsky2010, Madugani2012}, pressure tuning \cite{Ioppolo2007,Henze2011, Weigel2012, Martin2013,Henze2013}, electric field tuning, \cite{Ioppolo2009}, chemical etching \cite{White2005,Henze2013etch}, on-chip resistance heating \cite{Armani2004,Shainline2010}, and thermo-optic tuning \cite{Dong2009, Li2010,Watkins2012}. Each of these methods has its own distinct advantages and disadvantages and the choice of method depends ultimately on the final application.
We show that the microlasers fabricated using the glass-on-glass technique can be easily tuned over tens of GHz by one of two methods. The first method is unique to our devices and is applicable to the capillary structure; it relies on thermal tuning of the lasing mode by flowing gas through the cavity. From this method we demonstrate the idea of gas flow sensing using the concept of a “hot cavity” anemometer. Measurements and characterisation of the system as a gas flow sensor are presented. The second method relies on strain tuning and can be applied to both the tapered fibre and the capillary lasers.

\section*{Methods}
This method of microlaser fabrication can be applied to an optical fibre or capillary and is the same in both cases.  First, we tapered a microcapillary with an outer diameter (OD) of 350 $\mu$m and an inner diameter (ID) of 250 $\mu$m, using a CO$_2$ laser, to a uniform waist with an OD of $\sim$ 80 $\mu$m. Next, a glass wire was drawn from a bulk piece of  doped phosphate glass (Kigre). The diameter of the doped glass wire was typically a few tens of microns. The doped glass wire was then placed on top of and in contact with the microcapillary and the CO$_2$ laser beams were applied again. The CO$_2$ laser power was increased until the doped glass flowed onto the capillary, as shown in Fig. 1(a). At this point the doped glass wire was removed and the CO$_2$ laser power was controlled to allow the remaining doped glass to flow around the capillary. When the doped glass completely covered the capillary the CO$_2$ laser was turned off.  The resulting geometry was a hollow microbottle-shaped resonator. The diameter of the doped glass bottle was 170 $\mu$m and the thickness of the doped glass at the equator was 40 $\mu$m.  This thick layer means that any optical mode in the bottle cannot interact with the silica or any material inside the capillary.

\begin{figure}[h]
\centering\includegraphics[width=14cm]{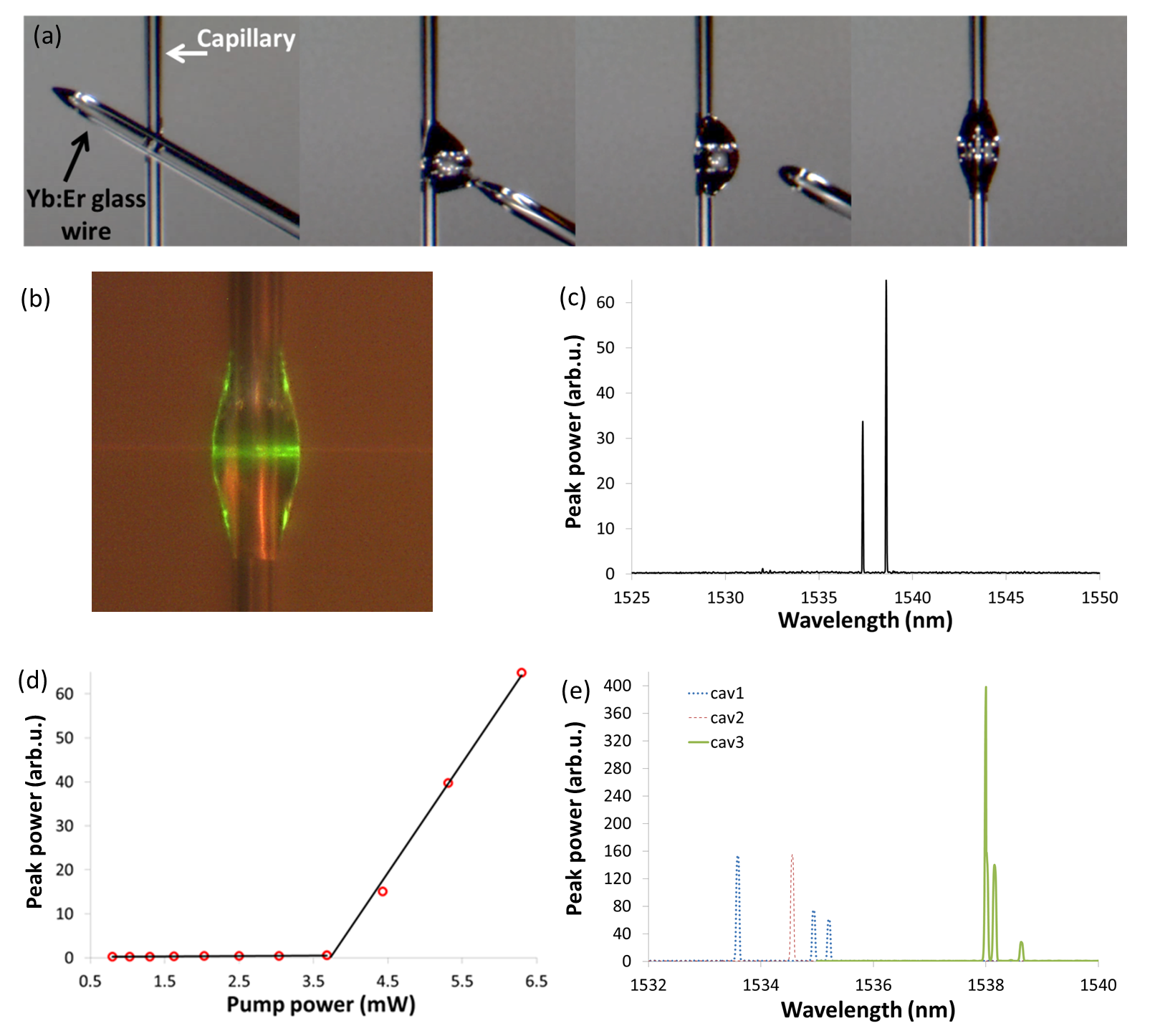}
\caption{(a) Fabrication steps for making a doped glass microbottle laser on a silica wire. The capillary diameter is 80 $\mu$m and the final diameter of the doped glass resonator is 170 $\mu$m. (b) Image of the resonator showing a WGM highlighted by green upconversion fluorescence. (c) Lasing spectrum from an Er:Yb doped bottle shaped resonator on a silica capillary. (d) Lasing threshold measurement. (e) Lasing spectra for three different microbottles (not that shown in parts (a) to (d)) on a single capillary with a diameter of 42 $\mu$m. The diameters of cav1, cav2 and cav3 are 120 $\mu$m, 170 $\mu$m and 155 $\mu$m, respectively}
\label{figure1}
\end{figure}

The capillary was glued onto a U-shaped holder and brought into contact with a tapered optical fibre for optical pumping. The tapered fibre had a diameter of $\sim$1 $\mu$m and was connected to a 980 nm diode laser. Whispering gallery modes were visible due to green upconversion fluorescence from the erbium ions, see Fig. 1(b). The output end of the tapered fibre was connected to an optical spectrum analyser (OSA). A typical lasing spectrum is shown in Fig. \ref{figure1}(c), with lasing peaks appearing between 1532 nm and 1540 nm. For a given microbottle the spectrum can be single mode or multimode depending on many factors such as the position of the tapered fibre along the capillary or the position of the capillary along the taper\cite{Yang2003}. To verify the lasing behaviour, a threshold measurement was taken, see Fig. \ref{figure1}(d). For this measurement the pump laser was not tuned to any particular WGM. The 980 nm pump laser used had a rated linewidth of 2 nm so it was assumed that a number of WGMs were simultaneously excited with no control over the coupling efficiency. As such, the pump power labled in Fig. \ref{figure1}(d) represents the pump power launched into the tapered fibre and does not represent the true pump power lasing threshold. Fig. \ref{figure1}(d) does, however, show a clear threshold for the peak output power.

By repeating the fabrication process at different points along a capillary it is possible to make a string of resonators in row. As a demonstration, three resonators with diameters of 120 $\mu$m, 170 $\mu$m and 155 $\mu$m  were made on the same capillary, which had an OD of 42 $\mu$m. Lasing was excited in each resonator in turn by simply moving the tapered fibre along the capillary to the next resonator. Lasing was collected from each resonator at the same coupling point on the tapered fibre and for the same pump power, see Fig. \ref{figure1}(e). 

\begin{figure}[htb]
\centering\includegraphics[width=17cm]{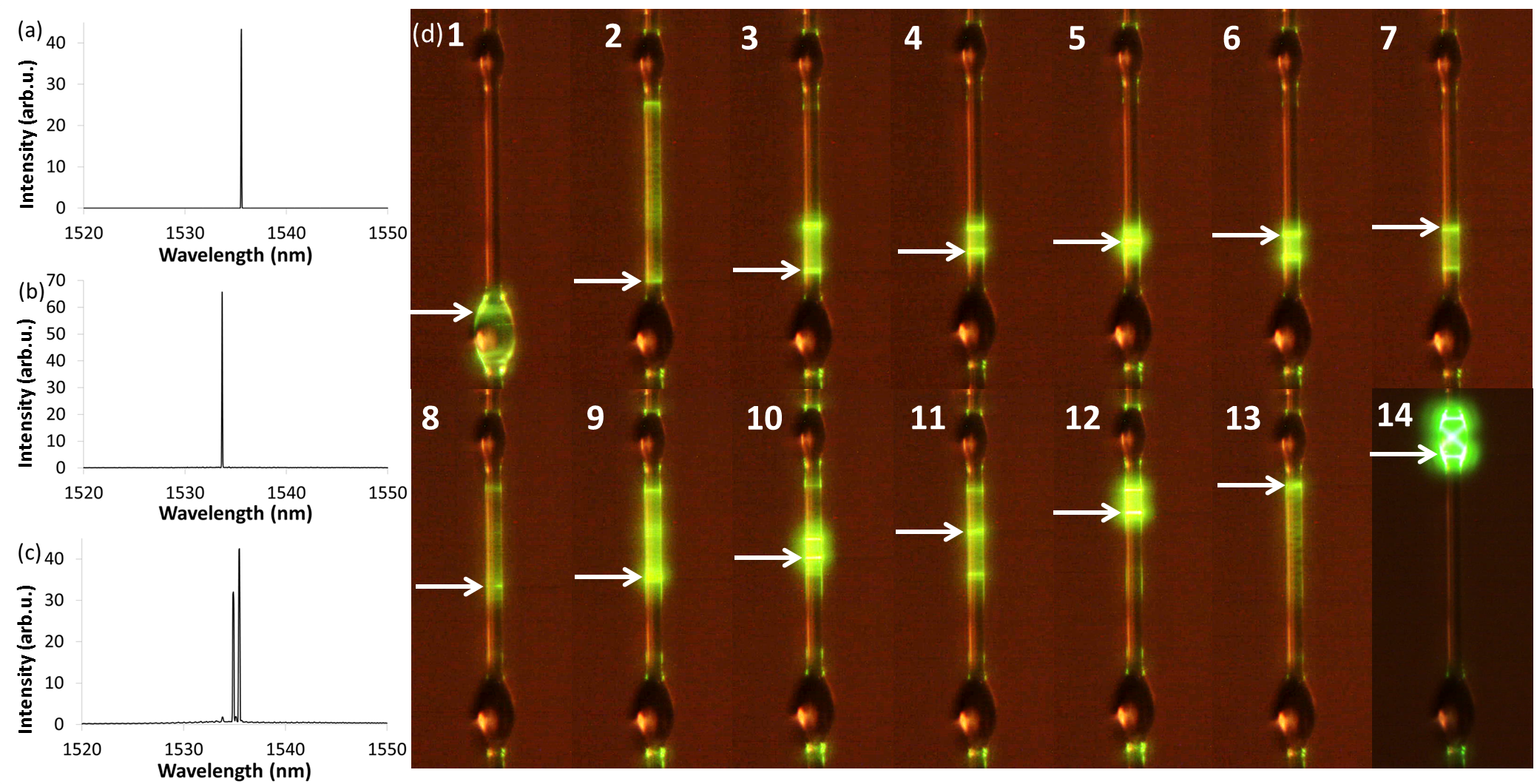}
\caption{(a-c) Lasing spectra from three of five resonators made on the same 20 $\mu$m fibre. The spectra correspond to bottle resonators with diameters of 42 $\mu$m (top), 60  $\mu$m (middle) and 160 $\mu$m (bottom). (d) The excitation of a continuous series of microresonators from thickness variations in a thin coating. The thin coating was formed between the 42 $\mu$m and 60 $\mu$m microbottles, seen at the top and bottom of each image with corresponding lasing spectra shown in (a) and (b).}
\label{figure2}
\end{figure}

The same fabrication steps were repeated for an optical fibre. In this case, a standard 780 nm single mode optical fibre  was tapered using a CO$_2$ laser to a diameter of 20 $\mu$m. The same erbium-doped glass was melted on the tapered fibre at five different points creating five separate cavities with diameters ranging from 42 $\mu$m to 232 $\mu$m. Fig. \ref{figure2}(a)-(c) shows the WGM lasing spectra taken for three of these resonators collected using the same pump power and position on the tapered fibre; note that the other two only showed fluorescence under these conditions.  

Apart from making a bottle-shaped or spherical resonator it was also possible to make a thin coating of the doped glass. This was achieved by heating the sphere of doped glass after it had been transferred to the fibre (or capillary). This additional heating caused the sphere to move along the fibre leaving behind a thin layer surrounding the fibre. The thickness of this layer is not constant so microcavities are formed by the small variations in the diameter, similar to SNAP structures\cite{Sumetsky2011}. Figure \ref{figure2}(d) shows such a thin layer spread out between two microlasers. The position of the excitation tapered fibre was moved along this thinly-coated region. The thickness of the doped layer was measured using an optical microscope and was determined to be around 1-2 $\mu$m. At each position a WGM spectrum was observed and, at some positions, lasing was achieved. In the future, a more accurate measurement of the diameter could be made form the variations in the WGM spectra at each position

\section*{Results}

\subsection*{Thermo-optical tuning and gas flow sensing}

\begin{figure}[htb]
\centering\includegraphics[width=11cm]{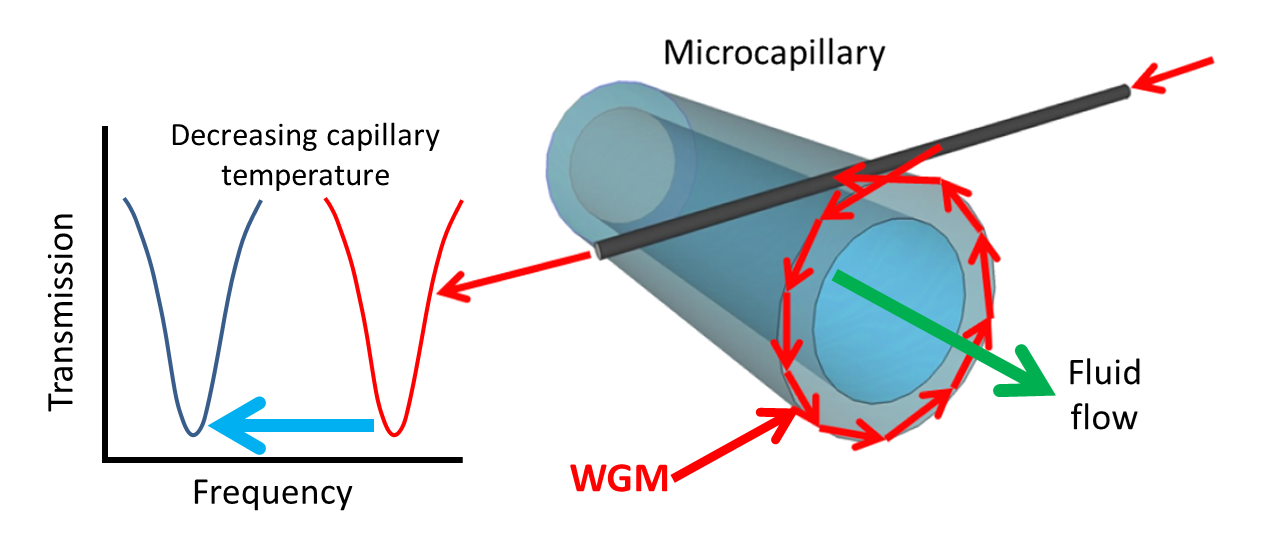}
\caption{Schematic of a WGM "hot cavity" anemometer. The excitation of the WGM is represented by the red arrows in the capillary wall. With sufficient absorbtion, the light in the WGM can locally heat the capillary. Fluid flowing through the capillary is represented by the green arrow. The flowing fluid removes the heat and shifts the WGMs  to higher frequencies, represented by the movement of the transmission dip on the left of the figure.}
\label{Figure3}
\end{figure}

Hollow whispering gallery resonators, such as microcapillaries \cite{Fan2007} and microbubbles \cite{Sumetsky2010, Yang2014,Ward2014}, have the unique feature that a material can fill or flow through their inner volume as already demonstrated for dye-filled microbubble lasers \cite{Lee2011}, capillaries \cite{Wu2009} and on-chip microfluidic channels \cite{Lee2011onchip}. The  light traveling in the WGM is partially absorbed by the glass and locally increases the temperature of the resonator's wall. When fluid flows through the resonator it removes some of the heat and reduces the temperature. This causes a blue shift in the frequency of the WGM. In this way the WGM modes can be tuned and the shift can be calibrated to represent the flow rate of the fluid, with larger flow rates giving larger blue shifts, see Fig. \ref{Figure3}.  

\noindent This is similar to the concept of all-optical, hot wire anemometry, a method which has been in use for some time.   Most of the reported devices are based on a SiO$_2$ optical waveguide, which is treated as a wire that absorbs light thereby creating heat \cite{Gao2011,Cheng2015,Cashdollar2005,Wang2013,Li2015}. The wire is placed into, or in contact with, a fluid flow channel. The flow of the fluid cools the wire and this changes the refractive index or the length of the wire.  Such changes can be read out optically using a fibre Braggg rating (FBG), for example\cite{Gao2011,Cheng2015,Cashdollar2005,Wang2013,Li2015}. Because SiO$_2$ is generally not a strong absorber, a significant amount of optical power is required to generate heat.  The process can be aided by the addition of a metal \cite{Li2015}. 
Most of the systems discussed above are not capable of measuring flow in arbitrarily small channels such as one would find in microfluidic systems, though recently a FBG hot wire anemometer was used to measure flow in such a system \cite{Li2015}. Although the sensor showed good sensitivity in the low flow rate regime, the device required hundreds of mW of pump power, the resolution was low, and good thermal contact needed to be maintained between the optical absorber and the flow channel. A hot cavity anemometer that can be incorporated onto a microcapillary automatically provides good thermal overlap between the sensor and the flow channel. For an Er:Yb doped resonator, significant heat is generated by pump absorption at 980 nm and a narrow linewidth lasing mode at 1535 nm is used to measure the thermo-optical shift of the cavity modes induced by the flow of fluid through the capillary. This device offers high resolution (by using the lasing modes) and high sensitivity (due to the high temperatures) with low pump powers.

\begin{figure}[ht]
\centering\includegraphics[width=16cm]{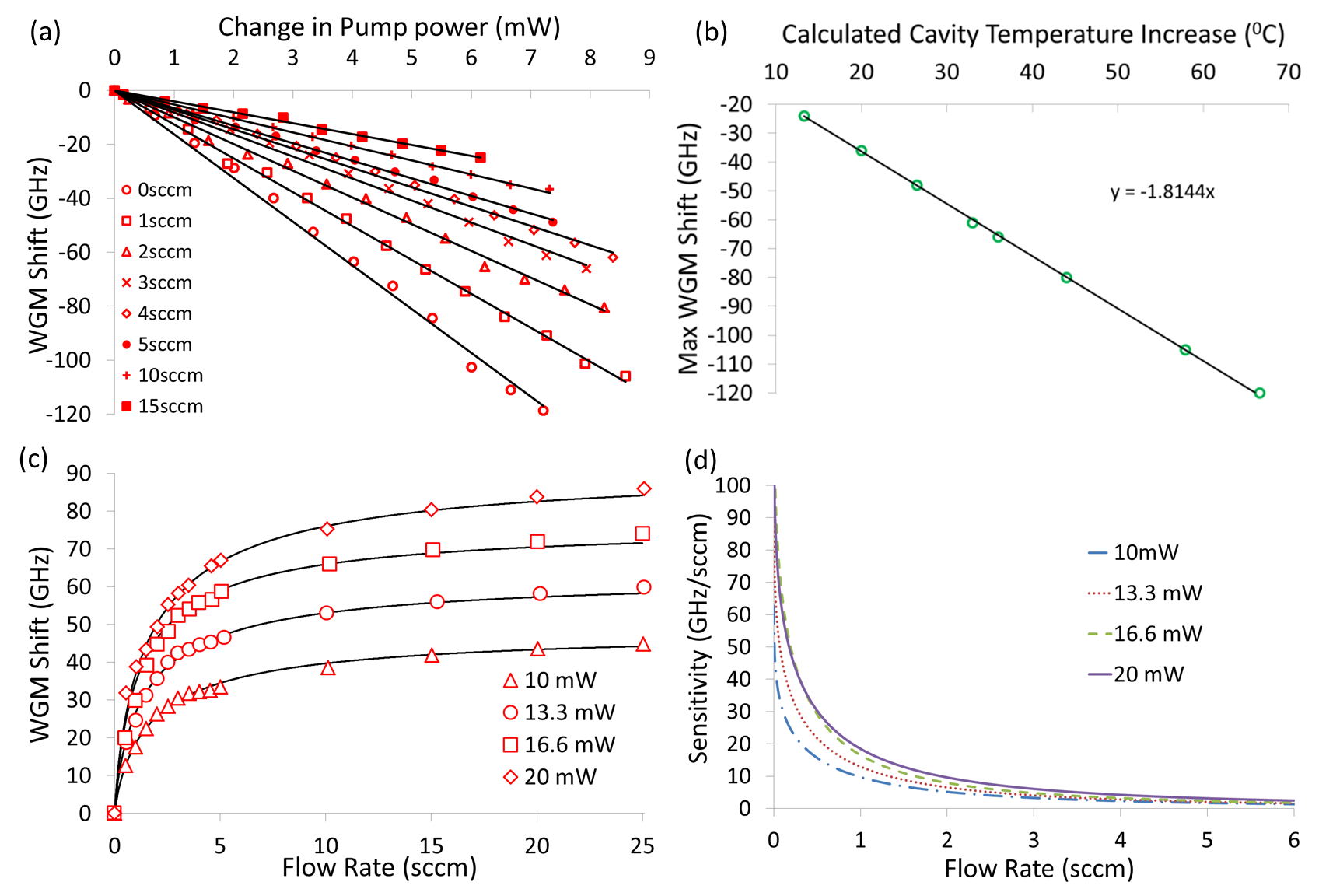}
\caption{(a) The shift rate of the 1535 nm lasing WGM as function of input pump power for different gas flow rates. (b) The calculated increase in cavity temperature and corresponding WGM shifts. (c) The shift of the 1535 nm lasing WGM as function of the measured flow rate for different pump powers. (d) The sensitivity of the WGM shift rate as function of measured flow rate, calculated from the derivative of the fits in (c).}
\label{figure4}
\end{figure}

It is well known that there are a number of large non-radiative energy transitions in erbium-doped glass which can be accessed by pumping at 980 nm \cite{Cai2002, ward2007, ward2010}. These phonon transitions generate a significant amount of heat in the glass even for low pump powers; in fact up to 40\% of the optical pumper power can be converted to heat. By flowing a gas/fluid through the capillary this heat is partially removed and the WGMs shift at a rate determined by the thermo-optical behaviour ($dn/dT = -21\times 10^{-7} K ^{-1}$) and the thermal contraction ($dr/dT =114\times 10^{-7} K ^{-1}$) of the glass. Based on the coefficients given by the manufacturer \cite{kigre} the thermal shift rate of the WGMs should be around 0.0145 nm/K (or 1.9 GHz/K) at 1535 nm, where the shift is defined as
\begin{gather}
\Delta\lambda=(\alpha+\beta)\Delta T,
\label{eq:delam}
\end{gather}

\noindent and \textit{$\Delta$T} is the change in cavity temperature. The capillary with the microlaser shown in Figs. 1(a) and 1(b) was connected to a source of pressurised air via a pressure regulator. The output of the capillary was connected to a mass flow meter. The shift of the lasing peaks was recorded as the pump power was increased. This was done with no air flowing through the capillary for increasing flow rates, with the results  plotted in Fig. \ref{figure4}(a). For this particular cavity, the shift rate of the WGMs goes from -16.2 GHz/mW to -4.1GHz/mW for zero flow and 15 sccm, respectively.  Using Eq. 1 the change in temperature of the cavity for the maximum shift in each case can be estimated and is plotted in Fig. \ref{figure4}(b). With no gas flow the temperature increases by 66$^{\circ}$C, whereas with a maximum flow rate of 15 sccm the maximum temperature increase was reduced to 13$^{\circ}$C. Next, the input pump power was fixed and the input pressure to the capillary was increased while the positions of the WGMs in the transmitted signal were recorded using the optical spectrum analyser. As the flow rate increased the WGMs were observed to shift towards shorter wavelengths due to the cooling effect of the air flow through the capillary, as discussed above.  This procedure was repeated for increasing pump powers and the results are plotted in Fig. \ref{figure4}(c). 

From the theory of optical hot wire anemometry \cite{Gao2011,Li2015}, the heat lost, \textit{H}, from a hot wire is related to the flow rate, $\nu$, by 
\begin{equation}
H=(A+B\nu^n)\Delta T_a
\label{eq:Hloss}
\end{equation}where \textit{A} and \textit{B} are empirical constants and \textit{n} is a fitting parameter (usually 0.5 for a simple hot wire).  Certain amount of pump power is used to generate the lasing signal so the cavity is already heated, thus the initial temperature of the cavity is not known. Even for modest pumping the initial cavity temperature can be easily greater than 100$^{\circ}$C \cite{Ward2008}. Therefore, we define \textit{$\Delta$T$_a$} = [\textit{T$_a$($\nu$)} - \textit{T$_a$($\nu$=0)}], i.e. the difference between the cavity temperature at zero flow and the cavity temperature at some flow rate, $\nu$. Based on the law of energy conservation the heat lost must equal the heat put in, therefore, in the case of the WGM resonator
\begin{equation}
H=I\eta(Q/Q_{abs}),
\label{eq:Hloss2}
\end{equation}

\noindent where \textit{I} is the input power, $\eta$ is the coupling efficiency, \textit{Q} is the quality factor of the cavity,  and \textit{Q$_{abs}$} is the absorption-limited cavity quality factor given by 
\begin{equation}
Q_{abs}=\frac{2\pi n_{eff}}{\sigma\lambda},
\label{eq:Qabs}
\end{equation}

\noindent where $\sigma$ is the absorption loss of the material, $\lambda$ is the wavelength and \textit{n$_{eff}$} is the effective refractive index. $\sigma$ was estimated from the glass material properties provided by the manufacturer \cite{kigre} using an absorbing ion concentration of 3$\times10^{21}$ ions/cm$^{3}$ and an absorption cross-section of 1.7$\times10^{20}$/cm$^{2}$. The calculated \textit{Q$_{abs}$} = 1.2$\times10^{6}$ and the loaded cavity $Q$ was assumed to be 100 times less than this.
To determine \textit{$\Delta$T$_a$}, the change in temperature for each pump power was determined from the maximum shift and Eq. 1. The WGM shift in terms of the gas flow rate can then be written as \cite{Gao2011,Li2015}
\begin{equation}
\delta\lambda=\lambda(\alpha+\beta)[H/(A+B\nu^n)-\Delta T_a].
\label{eq:delam2}
\end{equation}
The recorded WGM shifts were fitted using Eq. 5, see Fig. \ref{figure4}(c).  The 980 nm pump laser used in the experiment had a rated linewidth of 2 nm, therefore the coupling efficiency of the pump to a specific WGM was impossible to quantify. For fitting we assumed a coupling efficiency of 20$\%$. This value is justified by the fact that we observe about a 20-30$\%$ dip in transmitted power when the taper and microbottle make contact. The fitting parameter \textit{n} was set to 0.84. Using these parameters the corresponding fits agree with the observed shift. The sensitivities of the WGM shift rates are determined by differentiating the fits in Fig. \ref{figure4}(c) and are plotted in Fig. \ref{figure4}(d).

\subsubsection*{Strain tuning}

\begin{figure}[h]
\centering\includegraphics[width=10cm]{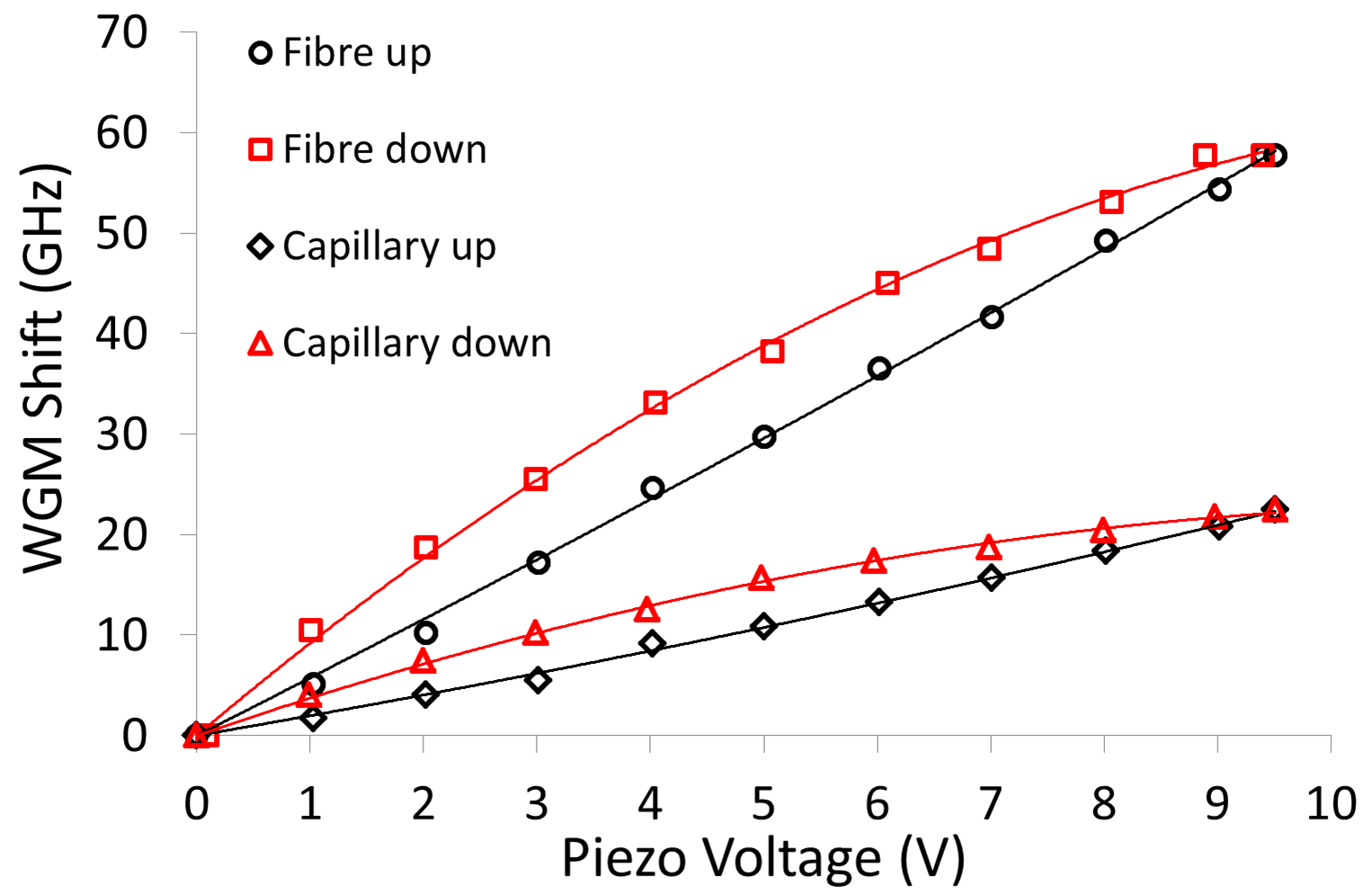}
\caption{Strain tuning of a 42 $\mu$m diameter microlaser on an optical fibre with a diameter of 20 $\mu$m and strain tuning of a 120 $\mu$m diameter microlaser on a capillary  with a diameter of 42 $\mu$m. Red curves: increasing piezo voltage, black curves: decreasing piezo voltage}
\label{figure5}
\end{figure}

Strain tuning of microresonators has been reported previously; it is an effective, fast and stable tuning method. Strain tuning of an erbium-doped microbottle laser was reported by P\"ollinger \textit{et al.} \cite{Pollinger2009}. However, etching using hydrofluoric acid was needed to access the core of an erbium-doped fibre which was subsequently melted by a CO$_2$ laser to form the bottle shape. In our work, no etching is required and a large number of microlasers with a wide range of sizes can be made quickly and easily. 
The 20 $\mu$m diameter fibre with the five microlasers, as described earlier, was held on a stage that allowed the U-shaped mount holding the fibre to be extended, thereby putting strain on the fibre. The stage was fitted with a piezo stack that provided 17 $\mu$m displacement. The 42 $\mu$m diameter microlaser was coupled to the tapered optical fibre and its lasing output was monitored while a voltage was applied to the piezo stack.  
The same procedure was applied to the 42 $\mu$m capillary supporting the three microlasers; in this case the 120 $\mu$m diameter microlaser was selected for tuning. The tuning curves for both microlasers are shown in Fig. \ref{figure5}.

In both cases the tuning curves experience a hysteresis due to the intrinsic behaviour of the piezo actuator. The tapered fibre laser showed a larger tuning of around 60 GHz, most likely due to its smaller diameter. However, a number of factors could influence the final tuning range, e.g. the diameter of the taper and the initial tension on the taper. For the capillary, the diameter and the wall thickness may also play a role. The taper or capillary and the doped glass microbottles are not a single structure so this may act to further reduce the final tuning range. Nevertheless, a significantly usable tuning range was achieved.

\section*{Discussion}
We have presented a  method for creating glass-on-glass structures, where one glass is melted and allowed to flow onto another tightly curved glass structure. This is achieved by the fact that the two glass have significantly different melting points; therefore, this method should also be applicable to other soft glasses. For example, high index glasses such as lead silicate or tellurite could be formed into small cavities on a tapered fibre with diameters around 20 $\mu$m. The number of resonators on a single fibre depends on the size of the resonator and length of the fibre.  So far we can place each cavity approximately 100 $\mu$m apart. One could envision a number of fibres mounted on  a chip  containing waveguides addressing each resonator.

The ability to make hollow, microbottle-shaped, doped-glass microlasers allows us to investigate a new method of thermo-optical tuning where the lasing modes can be tuned by simply flowing air through the cavity. The thickness of the capillary wall and a thick, doped-glass layer  negates any red shift of the optical mode induced by increases in internal pressure. We demonstrated that the mode shifts can be calibrated to represent the gas flow rate, thus creating an integrated, all-optical, flow sensor with low power and high resolution. The measurement of liquid flow is also possible with this setup and we have seen a water flow rate sensitivity of 1 GHz/(nL/sec) in a capillary with an ID of 100 $\mu$m. 
Early tests of strain tuning also show promising results and further studies should reveal the dependence of the tuning range on the fibre diameter, size of the resonator and wall thickness of the microcapillary.  In future work we plan to explore this method further using various soft glasses and structures to study the possibility of creating new and interesting glass-on-glass devices. For example, it maybe possible to flow melted glass into channels which have been pre-etched in silica glass.

\section*{Acknowledgments}

This work was funded by the Okinawa Institute of Science and Technology Graduate University.  

\section*{Author contributions statement}

J.M.W. conceived the experiment and did the data analysis; J.M.W. and Y.Y. conducted the measurements; S.N.C. supervised the project; all authors contributed to discussions on the project and manuscript preparation.  All authors reviewed the manuscript.

\section*{Additional information}
\textbf{Competing financial interests:} The authors declare no competing financial interests.

\end{document}